\documentclass[sigconf]{acmart}
\copyrightyear{2019}
\acmYear{2019}
\setcopyright{acmlicensed}
\acmConference[SIGIR '19] {42nd International ACM SIGIR Conference on Research and Development in Information Retrieval}{July 21--25, 2019}{Paris, France}
\acmBooktitle{42nd Int'l ACM SIGIR Conference on Research and Development in Information Retrieval (SIGIR '19), July 21--25, 2019, Paris, France}
\acmPrice{15.00}
\acmDOI{10.1145/3331184.3331274}
\acmISBN{978-1-4503-6172-9/19/07}

\begin{document}

%
\title{From Text to Sound: A Preliminary Study \\on Retrieving Sound Effects to Radio Stories}

%

\author{Songwei Ge}
\authornote{The work was done when the first author worked in Microsoft as an intern.}
\email{songweig@cs.cmu.edu}
\affiliation{%
  \institution{Carnegie Mellon University}
  \city{Pittsburgh}
  \state{Pennsylvania}
}

\author{Curtis Xuan}
\email{curtisxuan@ucla.edu}
\affiliation{%
  \institution{University of California, Los Angeles}
  \city{Los Angeles}
  \state{California}
}

\author{Ruihua Song}
\email{song.ruihua@microsoft.com}
\affiliation{%
  \institution{Microsoft XiaoIce}
  \city{Bejing}
  \country{China}
}

\author{Chao Zou}
\email{chazou@microsoft.com}
\affiliation{%
  \institution{Microsoft XiaoIce}
  \city{Bejing}
  \country{China}
}

\author{Wei Liu}
\email{wiliu@microsoft.com}
\affiliation{%
  \institution{Microsoft XiaoIce}
  \city{Bejing}
  \country{China}
}

\author{Jin Zhou}
\email{whitezj@vip.sina.com}
\affiliation{%
  \institution{Beijing Film Academy}
  \city{Bejing}
  \country{China}
}

\begin{abstract}
Sound effects play an essential role in producing high-quality radio stories but require enormous labor cost to add. In this paper, we address the problem of automatically adding sound effects to radio stories with a retrieval-based model. However, directly implementing a tag-based retrieval model leads to high false positives due to the ambiguity of story contents. To solve this problem, we introduce a retrieval-based framework hybridized with a semantic inference model which helps to achieve robust retrieval results. Our model relies on fine-designed features extracted from the context of candidate triggers. We collect two story dubbing datasets through crowdsourcing to analyze the setting of adding sound effects and to train and test our proposed methods. We further discuss the importance of each feature and introduce several heuristic rules for the trade-off between precision and recall. Together with the text-to-speech technology, our results reveal a promising automatic pipeline on producing high-quality radio stories. 
\end{abstract}


%
\maketitle

\section{Introduction}
\begin{figure}[t!]
    \centering
    \includegraphics[width=\linewidth]{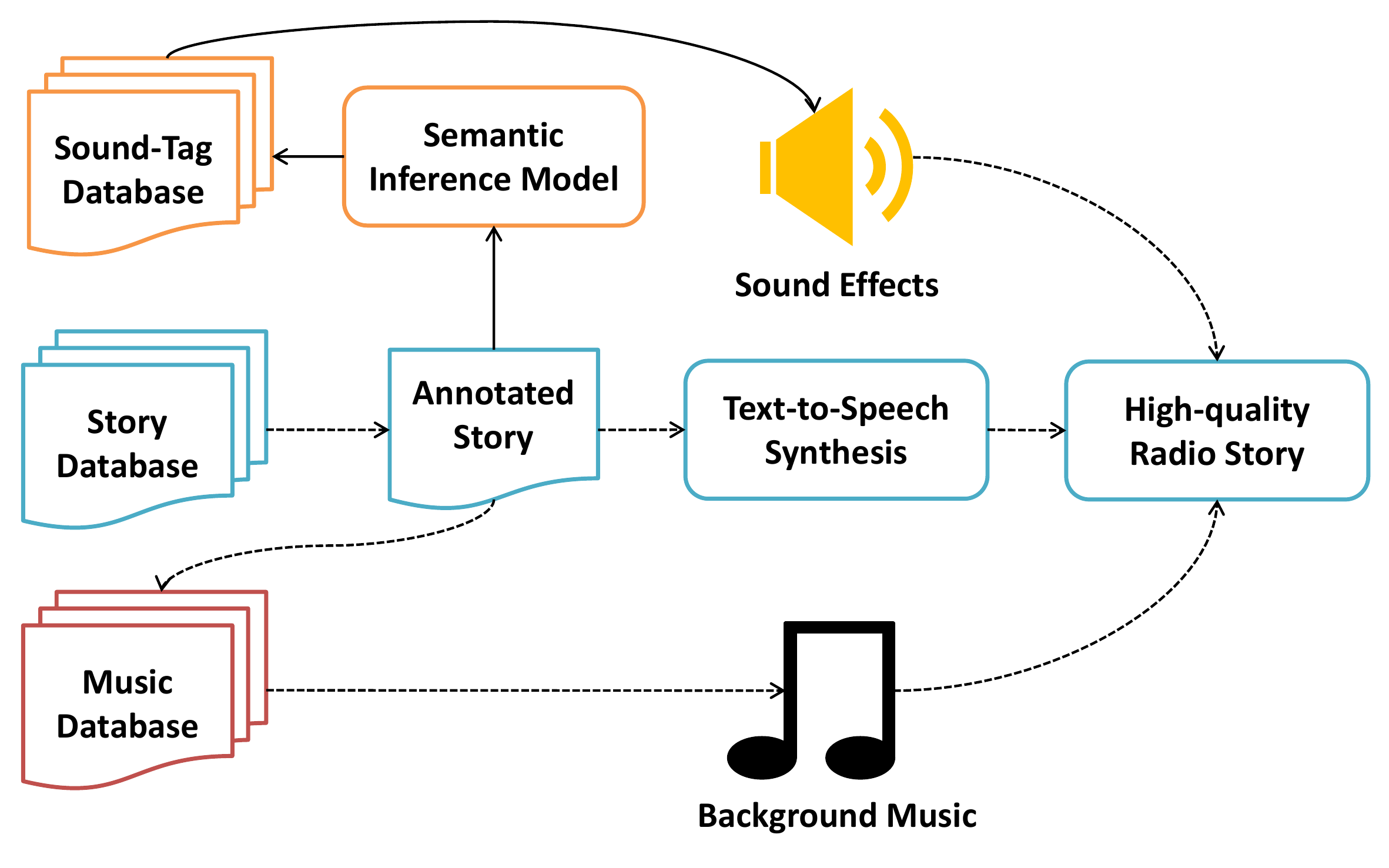}
    \caption{A sketch of radio story production pipeline. Note that the procedures indicated by the solid lines are considered in the task of this paper.}
    \label{fig:pipeline}
    \vspace{-0.3cm}
\end{figure}

Listening to stories is far more enjoyable when the listener can immerse him or herself into the story world. Radio stories provide a great acoustic experience of literature through emotional human reading. In addition to the human voices, the radio producer would manually add the sound effects that are implied by the stories in order to enhance the atmosphere. For example, certain sounds effects are expected to be played when the storyteller says "the wind is roaring" or "the Wolf Grandma knocked on the door". Though not seeing the actual scene in person, the listeners could reconstruct pseudo imagery in their minds with the help of these sounds. Such manually added sound effects have been found useful not only in helping listeners reconstruct the imagery but also in increasing their attention span~\cite{rodero2012see}. However, producing one radio story with sound effects takes onerous human work. With the mature development of text-to-speech technology~\cite{tabet2011speech} and its successful application on storytelling~\cite{doukhan2011prosodic,montano2013prosodic,montano2016role}, it is interesting to study the process of automatically adding sound effects to radio stories. Consequently, an end-to-end pipeline could be deployed for radio story generation as shown in the Figure~\ref{fig:pipeline}. In this paper, we explore the application of cross-modal retrieval methods on this problem, namely the inevitable difficulties in keyword search scenarios faced by a naive model and the potential solutions.

\begin{table}[!t]
  \begin{tabular}{p{3.8cm}||p{3.8cm}}
  	\toprule
    Sentences w/ Sound Effects & Sentences w/o Sound Effects \\
  	\midrule
    $\bullet$ Silly Bear runs over a rapid \textbf{river} through a wooden bridge with his honey pot. & $\bullet$ Among all the landscapes in this town, Mr. Cat loves the gold \textbf{beach} the most.\\
  	$\bullet$ A \textbf{thunder} suddenly booms over the heads. & $\bullet$ The rabbit doesn't \textbf{run} but looks around carefully.\\
	$\bullet$ The giant turtle swims to the distant of the \textbf{ocean}. & $\bullet$ The two brothers take a train towards the \textbf{forest}. \\
    $\bullet$ Dr.Tree shows up at the famous dancing \textbf{party}. & $\bullet$ John and his friends plan to hold a new year \textbf{party}.\\
    $\bullet$ In the \textbf{forest} there lives a family of bears. & $\bullet$ The woodpecker is well-known as the \textbf{forest} doctor.\\
    $\bullet$ He can't believe that he made it to the tumultuous \textbf{city} & $\bullet$ Little animals never get close to the dangerous \textbf{stream}. \\
  	\bottomrule
  \end{tabular}
  \caption{Examples of sentences with candidate triggers that should have sound effects and sentences that should not. The \textbf{bold} words are the candidate triggers returned by the retrieval model.}
  \label{tab:example}
  \vspace{-1.25cm}
\end{table}

Cross-modal retrieval has been studied in depth~\cite{wang2016comprehensive}, especially between text and images~\cite{kang2015learning, cao2017collective}. The basic retrieval models between text and sound also have a rich history, especially on the content-based audio retrieval~\cite{chechik2008large,turnbull2008semantic,lyon2010sound,roma2010ecological}. The quality of such models depends on the richness and accuracy of sound descriptions for training, which are designed deliberately and aligned perfectly with the sounds. In this paper, we focus on tag-based retrieval methods for the preliminary results of adding sound effects to radio stories. However, unexpected difficulties need to be addressed first under such a practical setting. To be specific, texts are often loosely related to the sounds and contain lots of noise, and simple keywords cannot be used consistently as reliable queries to retrieve sound effects.

Datasets of public sound effects with descriptive tags are common such as Adobe Audition Sound Effects~\footnote{https://offers.adobe.com/en/na/audition/offers/audition\_dlc/AdobeAuditionDLCSFX.html}, Freesound~\footnote{https://freesound.org/browse/tags/} and SoundBible~\footnote{http://soundbible.com/free-sound-effects-1.html}. We establish a sound-tag database with these resources and further enrich the tags to allow basic retrieval algorithm, such as BM25~\cite{robertson2009probabilistic}, to match the sound effects with stories. However, the retrieved results are prone to be unreliable. Specifically, not all of the tags in stories, which are called candidate triggers in this paper, indicate sound effects on a semantic level. For example, in our mind, the candidate trigger "forest" is related to a scene sound effect that may contain the voice of birds and insects. But in a story, the word "forest" could potentially be combined with other words and form a phrase such as "the king of forest", which is not describing a scene at all. In addition, even if the word "forest" is describing a scene, it might not be the scene where the story is taking place. For example, the word "forest" in "Little Piggy plans to go on a picnic in the forest" is describing forest as a scene, yet the current location is not the forest. Since the piggy has not gone to the forest yet, it would be unfit to play a sound here. More examples are displayed in Table \ref{tab:example}. Further analysis reveals that over 40\% triggers suggested by the retrieved results should not be added with sound effects. Such observation indicates that accepting the results returned by retrieval-based methods indiscriminately leads to high false positives. Therefore, in this paper, we introduce a semantic inference model to discriminate the necessity of adding sound effects based on information from the context.

 As illustrated by Table \ref{tab:example}, such a model needs to comprehend the context. Based on the observations in an annotated dataset from crowdsourcing, we introduce several useful features and attest their importance with ablation experiments. We also propose to use additional rules to control the trade-off between precision and recall. All these methods are evaluated on another crowdsourced dataset which contains candidate sentences and judgments from three labelers. The overall results show that our model, based on the semantic information, is effective in improving the robustness of adding sound effects. Further ablation results provide valuable insights on designing an efficient algorithm for this task.
 
\section{Problem Formulation}

Although we understand that the story text is sometimes connected to the sound, we do not precisely know when and how this connection happens. To better understand the elemental relationship between story and sound, we preliminarily annotate around 1300 radio stories, which are crawled from the web and with general public copyrights. During the process of annotation, we ask the human labelers to think about two key questions: 1. Are there any key words, i.e. candidate triggers, in the stories that might correspond to sounds? If so, what category of sound effects correspond to each trigger? 2. How confident do you think the sound effect does happen in the context of the story? The first question enables us to have a glimpse of the connections on the literal level, while the second one focuses on the semantic level. The human labelers are asked to span the candidate triggers and choose both the categories of the sounds and their levels of confidence ranging from 0 to 2. A larger number indicates a more reliable connection and 0 represents no semantic-level connection. We conduct a statistical analysis of the annotation data and report our results briefly as follows. 

\begin{table}[!t]
  \begin{tabular}{c||c}
  	\toprule
    Item & Statistic \\
  	\midrule
    Number of stories annotated & 1393\\
  	Number of candidate triggers & 8700 \\
	Number of confident candidate triggers & 6427 \\  
    Average number of candidate triggers & 6.25 \\
    Average number of confident candidate triggers & 4.61\\
    Std of candidate triggers & 6.20 \\
    Std of confident candidate triggers & 4.70 \\
  	\bottomrule
  \end{tabular}
  \caption{Overall statistics of the annotation.}
  \label{tab:annotation1}
  \vspace{-0.7cm}
\end{table}

There is a large number of possible places for adding sound effects in the stories: as shown in Table \ref{tab:annotation1}, around 6.25 key candidate triggers are found related to sound effects per story. As for the confidence of the connections, there are on average 4.61 candidate triggers labeled with confidence larger than 0, which implies that 1.64 candidate triggers in each story do not indicate real sound effects on the semantic level. Continuing our analysis, we divide text-triggered sound effects into four categories: action-triggered sound, scene-triggered sound, character-triggered sound, and onomatopoeia-triggered sound. As expected, the onomatopoeia-triggered sound has the highest confidence of 91.2\%, which suggests that most of the retrieved results are correct. However, when it comes to scene-triggered sound, only 57.9\% of the candidate triggers are labeled confident to represent a sound effect. Many non-ideal situations could happen, such as those listed in Table~\ref{tab:example}.

Preliminary results of keyword-based retrieval reveal that a great coverage rate can be achieved with the tag-sound database. Then the remaining question is that relying on the retrieval results will give rise to low precision. The cause behind this is that machines do not understand whether the scenes indicated by the words are truly happening in the story world. Suppose that we have a sentence $W_{1:n}$ and a known candidate trigger $W_{i:j}$ given by the retrieval model where $1\leq i\leq j\leq n$, our task is to determine whether $W_{i:j}$ is referring to an effective sound effect on the semantic level based on the context $W_{1:n}$. In our experiments, we will focus on the $W_{i:j}$ in scene-triggered sound effects, since it is the most difficult category.
\vspace{-0.8cm}

\section{Semantic Inference Model}

Based on the previous discussions, retrieving sound effects using only keywords is not sufficient to guarantee the robustness of results. Instead, the whole context should be considered when adding the sound effects to a story. In this section, we introduce a model with fine-designed features that can make inference about the necessity of adding sound effects based on contextual information.

Robust matching between texts and sounds requires not only the literal connection but also the semantic information from the context. To exploit such information, we design a number of features inspired by the first annotated dataset. The features we design can be categorized into three kinds: special words, part of speech and syntactic features which are elaborated as follows.

\textbf{Special Words:} We find that some special words convey important information about whether the sound effects are happening. For example, if a sentence contains the word which implies a subjunctive situation such as "plan", "like" and so on, the event that is associated with sound effects is often not happening. Such rule lends itself to not only subjunctive words but also action words, weather-related words, negative words, and time-based words. We use the counts of these special words or phrases in the sentences under each category as the several numerical features. To be specific, action words are associated with actions that often involve sounds, such as knocking, crying, yelling, etc. Weather words are those related to weather, such as raining, thunderstorm, wind, snow, etc. Negative words are associated with nullity, such as not, cannot, none, etc. Time-based words include past, now and future words, such as had, just, currently, since then, after, etc.

\textbf{Part of Speech (POS):} We find that the POS features play an important role in deciding the connections between sound and text. For example, candidate triggers that function as nouns in sentences more likely to be related to the real scene when compared with those that function as adjectives. We employ a one-hot encoding to represent the part of speech which the words serve as in the sentence. In addition to candidate triggers themselves, we also use the part of speech of the words before and after the candidate triggers to provide more complete semantic information.

\textbf{Syntactic:} We utilize the results of dependency parsing analysis on the candidate triggers to set up more features. For example, we can see that if a candidate trigger is a subject, object, or attribute to a verb, it is likely to indicate a sound. On the contrary, if it is attributed to a noun, it is likely not related to any sounds and instead plays a descriptive role in the sentence. We also employ a one-hot encoding to represent the different results of dependency parsing analysis on the candidate triggers. 

Though simple, we prove that these intuitive features produce safer results than plain retrieval results. In our experiments, we employ two popular machine learning approaches, SVM and XGBoost, as our basic classifier.

\section{Experiments}
In this section, we build a dataset and conduct experiments to validate the effectiveness of our proposed models.

\subsection{Setup}

Cross-modal retrieval relies highly on the precision of tags that describe the content of the sound effects. Therefore, one key step is to establish a sound-tag dataset with rich keywords. We take the scene-trigger sound effect as the example and collect the sound effects data under corresponding categories and their descriptions from a famous Chinese public sound effect website~\footnote{http://sc.chinaz.com/yinxiao/}. We pre-process the descriptions of sound effects by removing all the words except for verbs and nouns, and use them as the basic tags. Next, we extend these tags with similar words from both synonym tools~\footnote{https://github.com/huyingxi/Synonyms} and pretrained word embeddings space. Our preliminary dataset includes sound effects and multiple synonymous tags for each of them.

In addition to the sound-tag dataset, we need another radio story dataset with ground truth for training and evaluation. We select the 16 most common scene-triggered sound effects and use their tags as queries to retrieve the candidate sentences from 632 new stories. Note that the top 16 categories of the scene triggers account for around 83.88\% of all scene-triggered sound effects. Focusing on these categories helps alleviate the noise caused by those long-tailed sound effects. The 16 kinds of scenes are "forest", "mountain", "river", "sea", "rain", "wind", "thunder", "park", "party", "farm", "plaza", "prairie", "restaurant", "pool", "school" and "campfire". Finally, we produce 2069 candidate sentences that may connect to real sound effects. Then we label all of the candidate sentences through crowdsourcing. Note that the stories contained in the new dataset are completely different from those in the dataset where our inspiration came from in order to test the effectiveness of these features.

\begin{table}[!t]
  \begin{tabular}{c||c||c}
  	\toprule
    Category & Count & Ratio \\
  	\midrule
    No labelers marked to play sound & 1251 & 0.6046\\
	One labeler marked to play sound & 265 & 0.1281\\
	Two labelers marked to play sound & 210 & 0.1015\\
	All three labelers marked to play sound & 336 & 0.1624\\
  	\bottomrule
  \end{tabular}
  \caption{Distribution of sentences in terms of labels.}
  \label{tab:ablation}
  \vspace{-1.cm}
\end{table}

This time, the human labelers are given sentences with highlighted candidate triggers, and are asked to determine whether the event-triggering-sounds should be played. Each sentence is labeled three times independently by three different labelers, and only the consistent labels are used in the later experiments. Table 3 shows the labelers' results. Since the dataset is relatively small, it is important to guarantee the correctness of the labels for the sentences. Therefore, we used only the sentences all three labelers agreed on. For the purpose of balancing the positive and negative labels, we sample 336 sentences marked as no-play-sound arbitrarily from the 1251 sentences. Together with 336 play-sound sentences, totalling 672 sentences are used for training and evaluation. 

As for evaluation, we use 5-fold cross-validation and report the four model evaluation metrics: precision, recall, accuracy, and F1. Among these metrics, we care most about precision due to the goal of achieving robust retrieval results.


\subsection{Results}

In our experiments, We notice that the overall results of SVM and XGBoost are very similar. Thus, we simply present the results obtained from SVM in Table~\ref{tab:ablation}. Note that the table includes results derived from multiple experiments using different variations of features. In doing so, we conduct an ablation experiment to identify the most significant feature. Judging from the table, all of the feature combinations achieve precision higher than 50\%. Since we use half-and-half positive and negative test data, our model is verified to obtain more robust results than a simple retrieval model. As for the comparison between different feature categories, the category of special words is generally the most important for correct predictions. Not including it drops the precision by 8\%, accuracy by 7\%, and F1 by 5\%, while omitting the other two categories has less significant changes in the results. Out of all features of special words, the presence of action words in a sentence is the most essential, as omitting this feature causes a drop of 4\%. Omitting other special word features individually only causes a mere drop of around 1\% on average, so we do not display these results. It is also surprising that including the presence of now (part of time-based category) words actually worsens the predictions in every aspect by a considerable amount. One potential explanation of such observation is that the signal contained by now words are hard to generalize in newly seen scenarios, causing a discrepancy in distribution. The POS category seems to have the strongest effect on recall, and the syntactic feature has overall the weakest effect on results.

\begin{table}[!t]
  \begin{tabular}{p{2.2cm}||p{1.2cm}||p{0.9cm}||p{1.2cm}||p{0.8cm}}
  	\toprule
    Feature Excluded & Precision & Recall & Accuracy & F1 \\
  	\midrule
    None & 0.7022 & 0.7718 & 0.7195 & 0.7313\\
    Special Words & 0.6213 & 0.7758 & 0.6501 & 0.6877\\
	Action Words  & 0.6614 & \textbf{0.8061} & 0.6960 & 0.7246\\
	\textbf{Now Words} & \textbf{0.7167} & 0.8033 & \textbf{0.7410} & \textbf{0.7547}\\
	 POS & 0.6986 & 0.6990 & 0.6980 & 0.6961 \\
     Syntactic & 0.6876 & 0.7572 & 0.7046 & 0.7173\\
  	\bottomrule
  \end{tabular}
  \caption{Results of ablation experiment using SVM.}
  \label{tab:ablation}
  \vspace{-1.cm}
\end{table}

\section{Discussion on Additional Rules}

Some features are either hard to encode or already very strong to make the inference on their own, so we use the heuristic of combining features-based classification with additional rules. For this method, we use the combined features that give the best result, which is all but the feature that tracks the presence of now words. Then, we manually add a few rules: no sound should be played if 1. the scene words are in quotations or after colons; 2. the scene words appear after phrases including but not limited to "as if", "as though", "like", indicating a simile or metaphor. Using a combination of features and rules, we expect to see more robust results.

Interpreting from the results in Table~\ref{tab:rules}, only precision improved by a small amount, while results in other metrics dropped. The experiment shows that solely applying feature-based classification yields the overall best results. However, for the real-life application of adding sound effects to radio stories, this method would be valuable since we care more about whether the added sounds are appropriate than whether we miss opportunities to add sounds. 

\begin{table}[!t]
  \begin{tabular}{p{1.7cm}||p{1.2cm}||p{1.2cm}||p{1.2cm}||p{0.8cm}}
  	\toprule
    Strategy & Precision & Recall & Accuracy & F1 \\
  	\midrule
	w/o rules & 0.7167 & \textbf{0.8033} & \textbf{0.7410} & \textbf{0.7547}\\
    w/ rules & \textbf{0.7544} & 0.6337 & 0.7114 & 0.6859\\
  	\bottomrule
  \end{tabular}
  \caption{Results with or without additional rules.}
  \label{tab:rules}
  \vspace{-0.6cm}
\end{table}

\section{Conclusion and Future Work}
The application of cross-modal retrieval models between texts and sound effects is less studied. In this paper, we focus on the task of retrieving sound effect to radio stories. However, we find that naively using keywords-based retrieval is not good enough because of the ambiguity of story content. Therefore, we analyze the potential to utilize a semantic inference model which helps to improve the precision of the retrieval. The results on an annotated dataset using crowdsourcing attest that our model successfully decreases the false positive rate. Further ablation results reveal the importance of each feature which might be useful in improving the efficiency of feature engineering. We also propose to use heuristic rules to further increase the precision which in turn undermines the recall and lead to a trade-off problem. As for future work, we suppose that it is important to study the signals contained in a larger range than one single sentence. One promising way is to apply the neural network to exploit the information on the sentence level and circumvent onerous feature engineering work if more data is available.

\bibliography{main.bib}


\begin{thebibliography}{13}


\ifx \showCODEN    \undefined \def \showCODEN     #1{\unskip}     \fi
\ifx \showDOI      \undefined \def \showDOI       #1{#1}\fi
\ifx \showISBNx    \undefined \def \showISBNx     #1{\unskip}     \fi
\ifx \showISBNxiii \undefined \def \showISBNxiii  #1{\unskip}     \fi
\ifx \showISSN     \undefined \def \showISSN      #1{\unskip}     \fi
\ifx \showLCCN     \undefined \def \showLCCN      #1{\unskip}     \fi
\ifx \shownote     \undefined \def \shownote      #1{#1}          \fi
\ifx \showarticletitle \undefined \def \showarticletitle #1{#1}   \fi
\ifx \showURL      \undefined \def \showURL       {\relax}        \fi
\providecommand\bibfield[2]{#2}
\providecommand\bibinfo[2]{#2}
\providecommand\natexlab[1]{#1}
\providecommand\showeprint[2][]{arXiv:#2}

\bibitem[\protect\citeauthoryear{Cao, Long, Wang, and Liu}{Cao
  et~al\mbox{.}}{2017}]%
        {cao2017collective}
\bibfield{author}{\bibinfo{person}{Yue Cao}, \bibinfo{person}{Mingsheng Long},
  \bibinfo{person}{Jianmin Wang}, {and} \bibinfo{person}{Shichen Liu}.}
  \bibinfo{year}{2017}\natexlab{}.
\newblock \showarticletitle{Collective Deep Quantization for Efficient
  Cross-Modal Retrieval.}. In \bibinfo{booktitle}{\emph{AAAI}},
  Vol.~\bibinfo{volume}{1}. \bibinfo{pages}{5}.
\newblock


\bibitem[\protect\citeauthoryear{Chechik, Ie, Rehn, Bengio, and Lyon}{Chechik
  et~al\mbox{.}}{2008}]%
        {chechik2008large}
\bibfield{author}{\bibinfo{person}{Gal Chechik}, \bibinfo{person}{Eugene Ie},
  \bibinfo{person}{Martin Rehn}, \bibinfo{person}{Samy Bengio}, {and}
  \bibinfo{person}{Dick Lyon}.} \bibinfo{year}{2008}\natexlab{}.
\newblock \showarticletitle{Large-scale content-based audio retrieval from text
  queries}. In \bibinfo{booktitle}{\emph{Proceedings of the 1st ACM
  international conference on Multimedia information retrieval}}. ACM,
  \bibinfo{pages}{105--112}.
\newblock


\bibitem[\protect\citeauthoryear{Doukhan, Rilliard, Rosset, Adda-Decker, and
  d'Alessandro}{Doukhan et~al\mbox{.}}{2011}]%
        {doukhan2011prosodic}
\bibfield{author}{\bibinfo{person}{David Doukhan}, \bibinfo{person}{Albert
  Rilliard}, \bibinfo{person}{Sophie Rosset}, \bibinfo{person}{Martine
  Adda-Decker}, {and} \bibinfo{person}{Christophe d'Alessandro}.}
  \bibinfo{year}{2011}\natexlab{}.
\newblock \showarticletitle{Prosodic analysis of a corpus of tales}. In
  \bibinfo{booktitle}{\emph{Twelfth Annual Conference of the International
  Speech Communication Association}}.
\newblock


\bibitem[\protect\citeauthoryear{Kang, Xiang, Liao, Xu, and Pan}{Kang
  et~al\mbox{.}}{2015}]%
        {kang2015learning}
\bibfield{author}{\bibinfo{person}{Cuicui Kang}, \bibinfo{person}{Shiming
  Xiang}, \bibinfo{person}{Shengcai Liao}, \bibinfo{person}{Changsheng Xu},
  {and} \bibinfo{person}{Chunhong Pan}.} \bibinfo{year}{2015}\natexlab{}.
\newblock \showarticletitle{Learning consistent feature representation for
  cross-modal multimedia retrieval}.
\newblock \bibinfo{journal}{\emph{IEEE Transactions on Multimedia}}
  \bibinfo{volume}{17}, \bibinfo{number}{3} (\bibinfo{year}{2015}),
  \bibinfo{pages}{370--381}.
\newblock


\bibitem[\protect\citeauthoryear{Lyon, Rehn, Bengio, Walters, and Chechik}{Lyon
  et~al\mbox{.}}{2010}]%
        {lyon2010sound}
\bibfield{author}{\bibinfo{person}{Richard~F Lyon}, \bibinfo{person}{Martin
  Rehn}, \bibinfo{person}{Samy Bengio}, \bibinfo{person}{Thomas~C Walters},
  {and} \bibinfo{person}{Gal Chechik}.} \bibinfo{year}{2010}\natexlab{}.
\newblock \showarticletitle{Sound retrieval and ranking using sparse auditory
  representations}.
\newblock \bibinfo{journal}{\emph{Neural computation}} \bibinfo{volume}{22},
  \bibinfo{number}{9} (\bibinfo{year}{2010}), \bibinfo{pages}{2390--2416}.
\newblock


\bibitem[\protect\citeauthoryear{Monta{\~n}o and Al{\'\i}as}{Monta{\~n}o and
  Al{\'\i}as}{2016}]%
        {montano2016role}
\bibfield{author}{\bibinfo{person}{Ra{\'u}l Monta{\~n}o} {and}
  \bibinfo{person}{Francesc Al{\'\i}as}.} \bibinfo{year}{2016}\natexlab{}.
\newblock \showarticletitle{The role of prosody and voice quality in indirect
  storytelling speech: Annotation methodology and expressive categories}.
\newblock \bibinfo{journal}{\emph{Speech Communication}}  \bibinfo{volume}{85}
  (\bibinfo{year}{2016}), \bibinfo{pages}{8--18}.
\newblock


\bibitem[\protect\citeauthoryear{Monta{\~n}o, Al{\'\i}as, and
  Ferrer}{Monta{\~n}o et~al\mbox{.}}{2013}]%
        {montano2013prosodic}
\bibfield{author}{\bibinfo{person}{Ra{\'u}l Monta{\~n}o},
  \bibinfo{person}{Francesc Al{\'\i}as}, {and} \bibinfo{person}{Josep Ferrer}.}
  \bibinfo{year}{2013}\natexlab{}.
\newblock \showarticletitle{Prosodic analysis of storytelling discourse modes
  and narrative situations oriented to Text-to-Speech synthesis}. In
  \bibinfo{booktitle}{\emph{Eighth ISCA Workshop on Speech Synthesis}}.
\newblock


\bibitem[\protect\citeauthoryear{Robertson, Zaragoza, et~al\mbox{.}}{Robertson
  et~al\mbox{.}}{2009}]%
        {robertson2009probabilistic}
\bibfield{author}{\bibinfo{person}{Stephen Robertson}, \bibinfo{person}{Hugo
  Zaragoza}, {et~al\mbox{.}}} \bibinfo{year}{2009}\natexlab{}.
\newblock \showarticletitle{The probabilistic relevance framework: BM25 and
  beyond}.
\newblock \bibinfo{journal}{\emph{FTIR}} \bibinfo{volume}{3},
  \bibinfo{number}{4} (\bibinfo{year}{2009}), \bibinfo{pages}{333--389}.
\newblock


\bibitem[\protect\citeauthoryear{Rodero}{Rodero}{2012}]%
        {rodero2012see}
\bibfield{author}{\bibinfo{person}{Emma Rodero}.}
  \bibinfo{year}{2012}\natexlab{}.
\newblock \showarticletitle{See it on a radio story: Sound Effects and Shots to
  Evoked Imagery and Attention on Audio Fiction}.
\newblock \bibinfo{journal}{\emph{Communication research}}
  \bibinfo{volume}{39}, \bibinfo{number}{4} (\bibinfo{year}{2012}).
\newblock


\bibitem[\protect\citeauthoryear{Roma, Janer, Kersten, Schirosa, Herrera, and
  Serra}{Roma et~al\mbox{.}}{2010}]%
        {roma2010ecological}
\bibfield{author}{\bibinfo{person}{Gerard Roma}, \bibinfo{person}{Jordi Janer},
  \bibinfo{person}{Stefan Kersten}, \bibinfo{person}{Mattia Schirosa},
  \bibinfo{person}{Perfecto Herrera}, {and} \bibinfo{person}{Xavier Serra}.}
  \bibinfo{year}{2010}\natexlab{}.
\newblock \showarticletitle{Ecological acoustics perspective for content-based
  retrieval of environmental sounds}.
\newblock \bibinfo{journal}{\emph{EURASIP Journal}}  \bibinfo{volume}{2010}
  (\bibinfo{year}{2010}), \bibinfo{pages}{7}.
\newblock


\bibitem[\protect\citeauthoryear{Tabet and Boughazi}{Tabet and
  Boughazi}{2011}]%
        {tabet2011speech}
\bibfield{author}{\bibinfo{person}{Youcef Tabet} {and} \bibinfo{person}{Mohamed
  Boughazi}.} \bibinfo{year}{2011}\natexlab{}.
\newblock \showarticletitle{Speech synthesis techniques. A survey}. In
  \bibinfo{booktitle}{\emph{WOSSPA 2011 7th International Workshop on}}. IEEE,
  \bibinfo{pages}{67--70}.
\newblock


\bibitem[\protect\citeauthoryear{Turnbull, Barrington, Torres, and
  Lanckriet}{Turnbull et~al\mbox{.}}{2008}]%
        {turnbull2008semantic}
\bibfield{author}{\bibinfo{person}{Douglas Turnbull}, \bibinfo{person}{Luke
  Barrington}, \bibinfo{person}{David Torres}, {and} \bibinfo{person}{Gert
  Lanckriet}.} \bibinfo{year}{2008}\natexlab{}.
\newblock \showarticletitle{Semantic annotation and retrieval of music and
  sound effects}.
\newblock \bibinfo{journal}{\emph{IEEE Transactions on Audio, Speech, and
  Language Processing}} \bibinfo{volume}{16}, \bibinfo{number}{2}
  (\bibinfo{year}{2008}), \bibinfo{pages}{467--476}.
\newblock


\bibitem[\protect\citeauthoryear{Wang, Yin, Wang, Wu, and Wang}{Wang
  et~al\mbox{.}}{2016}]%
        {wang2016comprehensive}
\bibfield{author}{\bibinfo{person}{Kaiye Wang}, \bibinfo{person}{Qiyue Yin},
  \bibinfo{person}{Wei Wang}, \bibinfo{person}{Shu Wu}, {and}
  \bibinfo{person}{Liang Wang}.} \bibinfo{year}{2016}\natexlab{}.
\newblock \showarticletitle{A comprehensive survey on cross-modal retrieval}.
\newblock \bibinfo{journal}{\emph{arXiv preprint :1607.06215}}
  (\bibinfo{year}{2016}).
\newblock


\end{thebibliography}
\bibliographystyle{ACM-Reference-Format.bst}
\end{document}